\documentclass{article}
\pdfoutput=1

\usepackage{amssymb}
\usepackage{color}
\usepackage[usenames,dvipsnames,svgnames,table]{xcolor}
\usepackage{soul}

\def\be{\begin{equation}}
\def\ee{\end{equation}}
\def\ba{\begin{eqnarray}}
\def\ea{\end{eqnarray}}
\newcommand\DM{{\rm DM}}
\newcommand\obs{{\rm obs}}
\newcommand\IGM{{\rm IGM}}
\newcommand\MW{{\rm MW}}
\usepackage{mathptmx}
\usepackage{indentfirst}
\usepackage{amsmath,amssymb,latexsym,bm}
\usepackage{graphicx}	
\usepackage{amsmath}

\usepackage{jcappub}

\begin{document}

\title{Reconstruction of Reionization History through Dispersion Measurements of Fast Radio Bursts}
\author[a]{Ji-Ping Dai}
\author[a]{Jun-Qing Xia}
\affiliation[a]{Department of Astronomy, Beijing Normal University, Beijing 100875, China}

\emailAdd{daijp@mail.bnu.edu.cn}
\emailAdd{xiajq@bnu.edu.cn}

\abstract{
In this paper, we study the evolution of the ionization fraction $x_e(z)$ during the epoch of reionization by using the dispersion measurements (DMs) of fast radio bursts (FRBs). Different from the previous studies, here we turn to consider the large-scale clustering information of observed DMs of FRB catalog, which only needs the rough redshift distribution, instead of the exact redshift information of each FRB. Firstly, we consider the instantaneous ``\texttt{tanh}'' model for $x_e(z)$ and find that including the auto-correlation information of the mock catalog, about $10^4$ FRBs with the intrinsic DM scatter of 100 $\rm pc/cm^3$ {spanning 20\% of all sky}, could significantly improve the constraint on the width $\Delta_z$ of the model, when comparing with that from the CMB data alone. The evolution shape of the ionization fraction will be tightly narrowed, namely the duration of the epoch of reionization has been shrunk, {$z_{\rm dur}<2.24$ (95\% C.L.)}. Furthermore, we also use another redshift-asymmetric reionization model and obtain that the FRB mock catalog could measure the  ionization fraction at $z=6$ precisely with the $1\sigma$ error {$\Delta x_e(z=6)=0.012$}, which means that the large-scale clustering information of observed DMs of FRB catalog is very sensitive to the ionization fraction of the end of reionization epoch. We conclude that the observation of high-redshift FRBs could be a complementary probe to study the reionization history in the future.}

\maketitle

\section{Introduction}

Recently, the observation of the Cosmic Microwave Background (CMB) has provided state-of-the-art measurements on the cosmological parameters \citep{Aghanim:2018eyx} and given the tight constraints on the optical depth $\tau=0.0544^{+0,0070}_{-0.0081}$ (1$\sigma$ C.L.). However, since the parameter $\tau$ is an integration of the free electron density along the line of sight, the exact detail of the epoch of reionization, during which the mostly neutral hydrogen gas translates into the highly ionized state by the softer ionizing photos from the first stars and primordial dwarf galaxies, still remains one of the least well-understood aspects in the modern cosmology (for a recent review see Ref. \citep{2016ASSL..423.....M}). The simplest approach to modelling the reionization history is a step-like instantaneous model proposed by Ref. \citep{Lewis:2008wr}, which assumes a priori that there is a vanishing ionized fraction at early times, and a value of unity at low redshifts.

Recently, since the first detection of fast radio burst \citep{Lorimer:2007qn}, it has been a hotspot in astrophysics. FRBs are millisecond radio transients at $\sim$GHz frequencies characterized by the excess dispersion measure with respect to the Galactic values. The localization of the repeating sources FRB 121102 \citep{Spitler:2016dmz} and FRB 180924 \citep{Bannister:2019iju} provides us a good reason to believe FRBs originate from cosmological distance. As we know, we usually use the CMB polarization measurements to constrain the reionization history. However, since the DM of FRBs is affected by the total column density of free electrons along each line of sight from its sources, the high-redshift FRBs can help us to constrain the reionization history.

However, because of the lack of understanding on the nature of FRBs, it is still unknown whether FRBs can be detected at high redshift. There are certain progenitor models connecting FRBs with young neutron stars produced from supernovae or gamma-ray bursts (GRBs) \citep{Zhang:2013lta, Connor:2015era, Cordes:2015fua, Metzger:2017wdz}. Since GRBs with high redshifts ($z\sim9.4$) have been detected \citep{Cucchiara:2011pj}, it is possible that some FRBs can be generated at high redshift within these scenarios. Many current and upcoming surveys make FRB detections as one of their leading scientific goals, such as Parkes \citep{Petroff:2016tcr}, 
CHIME \citep{Amiri:2018qsq}, 
SKA \citep{Fialkov:2017qoz}, and they will be able to detect $\sim 10^4$ FRBs per decade. Ref. \citep{Fialkov:2016fjb} showed that SKA has sufficient sensitivity to probe FRBs out to $z\sim 14$, which makes it possible to constrain the reionization history using FRBs.

There are already some attempts which try to use the DM of FRBs combined with their redshift information to constrain the reionization history \citep{Zheng:2014rpa, Fialkov:2016fjb}. However it is not easy to measure the redshift of each FRB sample precisely. In this paper, we choose to use the large-scale clustering information to study the reionization history, which has been used to study the information of the host environment \citep{Shirasaki:2017otr}. Interestingly, we find that the auto-correlation information of DM from mock FRB samples, combining with the Planck 2018 measurements, could significantly improve the constraints on the evolution of the ionization fraction $x_e(z)$ during the epoch of reionization both in the instantaneous ``\texttt{tanh}'' model and redshift-asymmetric model. The structure of this paper is organized as follows. In Sec. \ref{sec:DMm}, we summarize the DM of FRBs and its possible clustering properties. We also present a theoretical model of the DM auto-correlation. In Sec. \ref{sec:md}, we show the parameterizations we adopted for the ionization fraction, which describe the reionization history as a function of redshift. In Sec. \ref{sec:data}, we then present the mock DM angular power spectra (APS) and the Planck 2018 measurements we included in this paper. Sec. \ref{sec:res} presents results based on the mock DM spectra and the Planck 2018 measurements by considering different reionization models. In particular, we derive limits on the reionization duration. Finally, we present conclusions in Sec. \ref{sec:con}.

\section{Large scale clustering of DM}
\label{sec:DMm}
\subsection{Properties of DM from FRBs}
The observed dispersion measure $\DM_{\obs}$ consists of the contributions from the intergalactic medium (IGM) $\DM_{\IGM}$, the FRB host galaxies $\DM_{\rm host}$ and the Milky Way $\DM_{\MW}$. Here, we assume that $\DM_{\MW}$ from each direction is already determined by Galactic pulsar observations \citep{Taylor:1993my}, and can be subtracted from $\DM_{\obs}$. In the following, we only focus on the extragalactic DM field.

$\DM_{\IGM}$ from a fixed source redshift $z_s$ is given by
\be
\mathrm{DM}_{\mathrm{IGM}}\left(\vec{\theta}, z_{s}\right)=\int_{0}^{z_{s}} \frac{\mathrm{d} z}{H(z)} \frac{n_{e}(\vec{\theta}, z)}{(1+z)^{2}}~,
\ee
where $n_e(\vec{\theta}, z)$ represents the number density of free electrons at redshift $z$. {Here we have used unit which $c=1$.} Relating the free electron density to the free electron fraction $x_e$, we can write
\be
n_e(\vec{\theta}, z) = \frac{x_e \rho_b}{m_p}{f_{\rm IGM}}f_{e}(z)~,
\ee
where $\rho_b$ is the baryon {mass} density, $m_p$ is the proton mass. {We should notice that $x_e$ is defined as the number of free electrons \emph{produced from hydrogen} per hydrogen atom, which will not be greater than 1 at any redshifts, and $f_{e}(z)$ can be written as}
{\be
f_{e}(z)= (1-Y)+ \frac{Y}{4x_e}\left[x_e+\frac{1}{2}\left(1+{\tanh}\left(\frac{3.5-z}{0.5}\right)\right)\right]~,
\ee}
where $Y\simeq0.24$ is the mass fraction of helium, the terms in the bracket account for the doubly ionized helium. The first ionization of helium is assumed to happen at the same time as hydrogen reionization, and the full reionization of helium happens fairly sharply at $z=3.5$ \citep{Becker:2010cu}. $f_{\rm IGM}$ is the fraction of free electrons in the intergalactic medium. In our analysis, we set $f_{\rm IGM} = 1$ {since there are fewer massive halos and the majority of baryons are contained
within the IGM  during reionization.}

Therefore, $\DM_{\IGM}$ at $z_s$ can be re-written as
\be
\mathrm{DM}_{\mathrm{IGM}}\left(\vec{\theta}, z_{s}\right)=\frac{ \bar\rho_{b,0}}{m_p}\int_{0}^{z_{s}}\mathrm{d} z  \frac{{(1+z)}}{H(z)} f_{e}(z)X(\vec{\theta}, z)~.
\ee
Here we use the shorthand notation $X(\vec{\theta}, z)=x_e(1+\delta_b)$, where $\delta_b$ is the baryon density perturbation and $\bar\rho_{b,0}$ is the average baryon mass density at present time.  Next, we preform the 2D spherical projection, which means we only need the {normalized number distribution} of FRB catalog $n(z)$ rather than the precise redshift information of each sample. The average $\DM_{\IGM}$ for an angular position $\vec{\theta}$ can be written as
\be
\label{eq:IGM}
{\DM}_{{\IGM}}(\vec{\theta}) = \int_0^{\infty} {{\rm d} z} W_{{\DM},{\IGM}}(z) X(\vec{\theta}, z)~,
\ee
where the window function is
\be
W_{{\DM},{\IGM}}(z)=f_{e}(z) \frac{ \bar\rho_{b,0}}{m_p}\frac{{(1+z)}}{H(z)} \int_z^{\infty} n(z) {\rm d}z~.
\ee
We also need to consider the average $\DM_{\rm host}$ for an angular position, following Ref. \citep{Shirasaki:2017otr}, we have
\be
\label{eq:host}
\mathrm{DM}_{\mathrm{host}}(\vec{\theta})=\int_{0}^{\infty} \mathrm{d} z W_{\mathrm{DM}, \mathrm{host}}(z)\left[1+\delta_{s}(\vec{\theta}, z)\right],
\ee
where $W_{\mathrm{DM}, \text { host }}(z)=\bar{\tau}_{e}(z) n(z)$, $\delta_s$ is the FRB number density perturbation and {$\bar{\tau}_{e}(z)$ represents the observed mean DM from host galaxies at redshift $z$ of the Earth observer}. Actually, the redshift dependence of $\bar{\tau}_{e}(z)$ contains the information of the environment of FRB sources, which is poorly known. In our paper, we assume the DM in host-frame $\bar{\tau}_{e,\rm host}(z)$ to be a constant, $\bar{\tau}_{e,\rm host}(z) = 100 \rm ~pc/cm^{3}$, which means $\bar{\tau}_{e}(z)=100/(1+z)\rm ~pc/cm^{3}$ in the Earth observer frame.

\subsection{DM auto-correlation}

\begin{figure}[tb]
	\centering
    \includegraphics[width=0.8\linewidth]{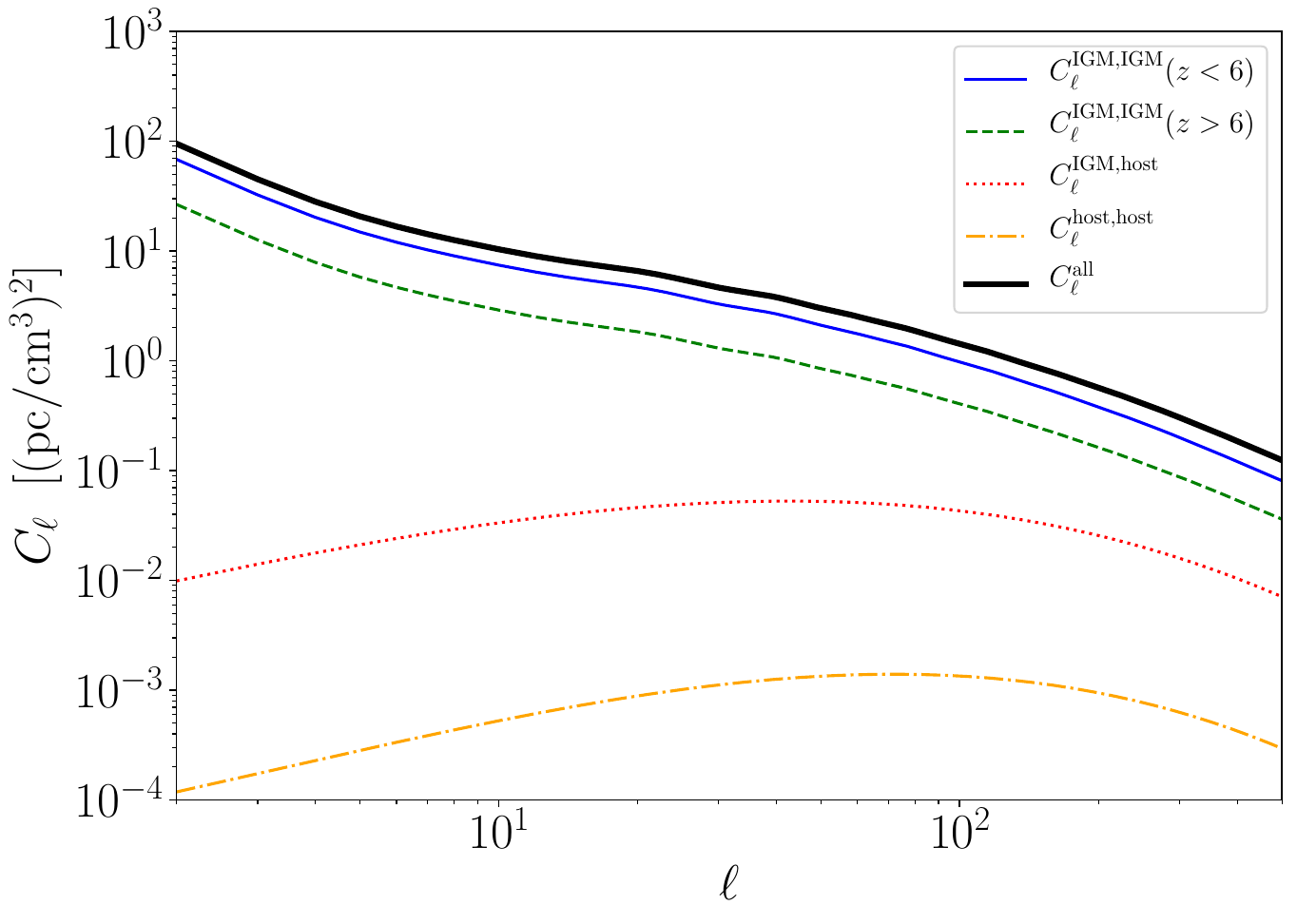}
	\caption{The angular power spectra of DM from auto-correlation of the IGM component using lower redshift samples (blue solid line) and higher redshift samples (green dashed line), the cross-correlation of the IGM and the host galaxy component (red dotted line), and auto-correlation of the host galaxy component (orange dotted-dashed line). We also plot their summation
(black solid line). Here we use the \texttt{tanh} instantaneous reionization model and the details can be found in text. }
	\label{fig:cl}
\end{figure}

For a FRB catalog, the auto-correlation power spectrum of DM: $C_\ell^{\DM,\DM}$ consists of three parts, $C^{\IGM,\IGM}_{\ell}$, $C^{\IGM,{\rm host}}_{\ell}$ and $C^{{\rm host},{\rm host}}_{\ell}$. Using Eq.(\ref{eq:IGM}), Eq(\ref{eq:host}) and the Limber approximation \citep{limber1953analysis}, we have
\be
\label{eq:ps}
C^{\mathrm{IGM},\mathrm{IGM}}_{\ell}=\int \mathrm{d} z W_{\mathrm{DM}, \mathrm{IGM}}^{2}(z) \frac{H(z)}{\chi^{2}(z)} P_{XX}\left( \frac{\ell+1/2}{\chi(z)},z \right)~,
\ee
\be
\begin{aligned}
C_{\ell}^{\text {IGM,host}}= 2\int \mathrm{d} z \, W_{\text {DM}, \text {IGM}}(z) W_{\text {DM}, \text {host}}(z) \frac{H(z)}{\chi^{2}(z)}  \times b_{\text {FRB}} P_{X m}\left(\frac{\ell+1 / 2}{\chi(z)}, z\right)~,
\end{aligned}
\ee
\be
C^{\text {host,host}}_{\ell}=\int \mathrm{d} z W_{\mathrm{DM}, \text {host}}^{2}(z) \frac{H(z)}{\chi^{2}(z)} b_{\rm FRB}^2P_{m}\left( \frac{\ell+1/2}{\chi(z)},z \right)~,
\ee
where $\chi(z)$ is the comoving distance, $P_m(k)$ is the matter power spectrum and the FRB bias $b_{\rm FRB} =\delta_s/\delta_m$. We assume FRBs form in dark matter halos, so $b_{\rm FRB}$ can be calculated from the halo bias using the fitting formula proposed by Ref. \citep{Tinker:2010my} using $N$-body simulations, and we set the halo mass $M=10^{13}h^{-1}M_{\odot}$ in our analysis. {Actually, $b_{\rm FRB}$ is not very sensitive to our final results since $C^{\mathrm{IGM},\mathrm{IGM}}_{\ell}$  dominates the signal as Fig. \ref{fig:cl} shows. Besides, the bias parameters can be obtained by the auto-correlation of FRB density field and
cross-correlation between FRB and galaxy density field (e.g. \citep{Shirasaki:2017otr})}.

The final step is explicating the three-dimensional power spectra of $P_{XX}(k)$ and $P_{Xm}(k)$. $P_{XX}(k)$ is a summation of two terms, corresponding to 1-bubble (1b) and 2-bubble (2b) contributions to the power spectrum of the ionized hydrogen \citep{Gruzinov:1998un, Wang:2005my, Mortonson:2006re, Dvorkin:2008tf,  Meerburg:2013dua}:
\be
P_{XX}(k)=P_{XX}^{1 \mathrm{b}}(k)+P_{XX}^{2 \mathrm{b}}(k).
\ee
The 1-bubble contribution to the three-dimensional power spectrum is given by
\be
\label{eq:pk1b}
P_{XX}^{1 \mathrm{b}}(k)=x_e(1-x_e)[F(k)+G(k)].
\ee
$F(k)$ and $G(k)$ can be written as
\be
F(k)=\frac{\int d R P(R)[V(R)]^{2}[W(k R)]^{2}}{\int d R P(R) V(R)},
\ee
\be
G(k)=\int \frac{d^{3} \vec{k}^{\prime}}{(2 \pi)^{3}} P_m\left(\left|\vec{k}-\vec{k}^{\prime}\right|\right) F\left(k^{\prime}\right),
\ee
where $V(R)=4\pi R^3/3$ is the volume of the bubble, $P(R)$ is the log-normal distribution which can be written as
\be
P(R)=\frac{1}{R} \frac{1}{\sqrt{2 \pi \sigma_{\ln R}^{2}}} {\rm e}^{-[\ln (R / \bar R)]^{2} /\left(2 \sigma_{\ln R}^{2}\right)},
\ee
here $\bar R$ and $\sigma_{\ln R}$ are the characteristic size and width of the distribution, respectively. $W(kR)$ is the Fourier transform of a real space top-hat window function,
\be
W(k R)=\frac{3}{(k R)^{3}}[\sin (k R)-k R \cos (k R)].
\ee
The 2-bubble contribution is given by
\be
\label{eq:pk2b}
P_{XX}^{2 \mathrm{b}}(k)=[(1-x_e)\ln(1-x_e)I(k)-x_e]^2P_m(k),
\ee
where the function $I(k)$ is
\be
I(k)=b \frac{\int d R P(R) V(R) W(k R)}{\int d R P(R) V(R)},
\ee
and $b$ is the bubble bias.

As for the cross power spectrum $P_{X, m}(k)$, it only has the 2-bubble term, which is given by \citep{Alvarez:2005sa, Tashiro:2008vg}
\be
P_{Xm}(k) = [-(1-x_e)\ln(1-x_e)I(k)+x_e] P_m(k).
\ee
{Clearly, at the end of reionization  when $x_e\simeq 1$, $P_{XX}(k)$ and $P_{Xm}(k)$ are equal to $P_{m}(k)$ as excepted.}
In our paper, we use the parameters $b=6$, ${\bar R}=5\rm Mpc$ and $\sigma_{\ln R}=\ln(2)$, which are adopted in Ref. \citep{Dvorkin:2008tf}.

{In Fig. \ref{fig:cl} we show the  auto-correlation angular power spectra using \texttt{tanh}  instantaneous reionization model (see details in Sec. \ref{sec:md}) and the mock FRB samples distribution (see details in Sec. \ref{sec:data}). The redshift of reionization $z_{\rm re}$ and the reionization width $\Delta_z$ are set to 7.7 and 0.5 respectively.
Here, we do not consider the host galaxies' cosmological evolution, since we find that the contribution from the IGM component dominates the signal. So  different assumption of the host galaxies does not significantly affect the final results.} {Additionally, the signal from higher redshifts ($z>6$) is subdominant comparing to the information from  lower redshifts, and the reason is that there are fewer higher redshift samples}.

\section{Reionization Models}
\label{sec:md}
Here, we adopt two reionization models. The first one is the most widely used parameterization which describes the epoch of reionization using a simple ``\texttt{tanh}'' function \citep{Lewis:2008wr}:
\be
{x}_{e}(z)=\frac{1}{2}(1-x_{e}^{\rm rec})\left[1+\tanh \left(\frac{y_{\mathrm{re}}-(1+z)^{3 / 2}}{\Delta_y}\right)\right]+x_{e}^{\rm rec},
\ee
where $y(z)=(1+z)^{3/2}$, $\Delta_y=1.5\sqrt{1+z_{\rm re}}\Delta_z$ and $x_{e}^{\rm rec}$ is the ionized fraction left over from recombination. Here $z_{\rm re}$ is defined as the redshift of reionization where $z_{\rm re}\equiv z_{50\%}$, at which $x_e=0.5$. To reconstruct the CMB spectra, this one-stage almost redshift-symmetric reionization transition is mainly determined by $z_{\rm re}$ and almost degenerate with the width $\Delta_z$. We expect to break this degeneracy with the DM measurements from FRBs.

Although the redshift-symmetric ``\texttt{tanh}'' model has been widely used, a redshift-asymmetric parameterization could be a better choice to describe the numerical simulations of the reionization flexibly \citep{Ahn:2012sb, Park:2013mv, Douspis:2015nca}, and is also suggested by the constraints from the ionizing background measurements \citep{Faisst:2014vra, Chornock:2014fva, Bouwens:2015vha, Ishigaki:2014cga, Robertson:2015uda}.
Here we adopt a redshift-asymmetric model described by two parameters: the ionization fraction at $z=6$, $x_{e,6}$ and the exponent $\alpha$, similar to Ref. \citep{Douspis:2015nca},
\be
\label{eq:as}
x_{e}(z)=\left\{\begin{array}{ll}
	1-\frac{(1-x_{e,6})(1+z)^3}{(1+6)^3} & \text { for } z\le6~, \\
	x_{e,6}{\rm e}^{\alpha(6-z)} & \text { for } z>6~.
\end{array}\right.
\ee

An important fact is that the observations of the Gunn-Peterson effect on high-redshift quasars showed that our Universe was almost fully reionized at $z\simeq 6$ \citep{Fan:2001vx}. However, there are still some measurements on Ly$\alpha$ Damping Wing of quasars pointed to $x_e<0.9$ at $2 \sigma$ C.L. between $z=6.24-6.42$ \citep{Schroeder:2012uy}, and other measurements on dark gaps in quasar spectra showed $x_e>0.89$ at $z=5.9$ ($1 \sigma$ C.L.) \citep{McGreer:2014qwa}.{ Actually, many works using different methods have been done to explore the reionization history with high-redshift quasars,  gamma-ray bursts or Ly$\alpha$ emitting galaxies \citep{Malhotra:2004ef, Fan:2005es,Kashikawa:2006pb, Totani:2005ng,McQuinn:2007dy, Gallerani:2007ty, Gallerani:2007nb, Ota:2007nx, Ouchi:2010wd,Mortlock:2011va,Ono:2011ew, Caruana:2013qua, Pentericci:2014nia,Schenker:2014tda,Tilvi:2014oia,Mesinger:2014mqa,Sobacchi:2015gpa,Greig:2016vpu, Zheng:2017rvy}. Despite these efforts, there is not a unified conclusion,
due to the uncertainties and degeneracies in modeling high-redshift astrophysics of the observational data.
To be conservative, in this paper, we consider $x_e(z=6)>0.9$ as a prior in the calculations to constrain the reionization history.}

\section{Data and likelihood}
\label{sec:data}

We perform a global fitting analysis of cosmological parameters, using the public \textsc{CosmoMC} package \citep{Lewis:2002ah}, a Markov Chain Monte Carlo code. We assume purely adiabatic initial conditions and a $\Lambda$CDM universe. Our parameterization is thus: $P\equiv \{\Omega_{b} h^{2}, \Omega_{c} h^{2}, \Theta_{s}, n_{s}, A_{s}, P_{\rm re}\}$, where $\Omega_{b} h^{2}$ and $\Omega_{c} h^{2}$ are the baryon and cold dark matter physical density, $\Theta_s$ is the angular size of the sound horizon at decoupling, $n_s$ and $A_s$ are the spectral index and amplitude of the primordial power spectrum, and {$P_{\rm re}$ are $\{ z_{\rm re}, \Delta_z\}$ and $\{x_{e,6}, \alpha\}$ for the two reionization models, respectively.}

For the dataset, we consider the measurements of CMB temperature and polarization anisotropy from the Planck 2018 legacy data release, which provide the utmost observations on temperature and polarization information from the last scattering surface. We use the combination of the \texttt{Plik} likelihood using $TT$, $TE$ and $EE$ spectra at $\ell\ge30$, the low-$\ell$ ($\ell=2\sim29$) temperature \texttt{Commander} likelihood and the \texttt{SimAll} $EE$ likelihood, which is the dataset labeled as TT,TE,EE+lowE in Ref. \citep{Aghanim:2018eyx}.

As for the mock FRB measurements, we do not need the precise redshift information of each sample which is hard to obtain. Instead, we only need the observed FRB distribution $n(z)$ {which can be estimated from observed DMs. What is more, the redshift distribution of FRBs also can be estimated from an empirical relation in the era of Square Kilometre Array (SKA) \citep{Hashimoto2021}, which can give us a complementary check}. Here we adopt the form used in Ref. \citep{Linder:2020aru}
\be
n(z)\sim z^3 {\rm e} ^{-z}.
\ee
The distribution reaches its peak at $z\sim3$ and there are $\sim$15\%  high-redshift ($z>6$) samples, which we expect can be achieved by future surveys \citep{Hashimoto:2020dud}. Furthermore, we calculate the shot noise induced by the intrinsic scatter of DM around host galaxies \citep{Shirasaki:2017otr, Reischke:2020cgd, Takahashi:2021}: $N^{\DM,\DM}_{\ell,\rm SN} = {4\pi}f_{\rm sky}\sigma^2_{\rm host}/{\mathcal{N}}$, and set the sky coverage $f_{\rm sky}=0.2$, the intrinsic scatter of DM around host galaxies $\sigma_{\rm host}=100 \rm~pc/cm^{3}$, and the total number of FRBs $\mathcal{N}=10000$.

\begin{figure}[tb]
	\centering
    \includegraphics[width=1\linewidth]{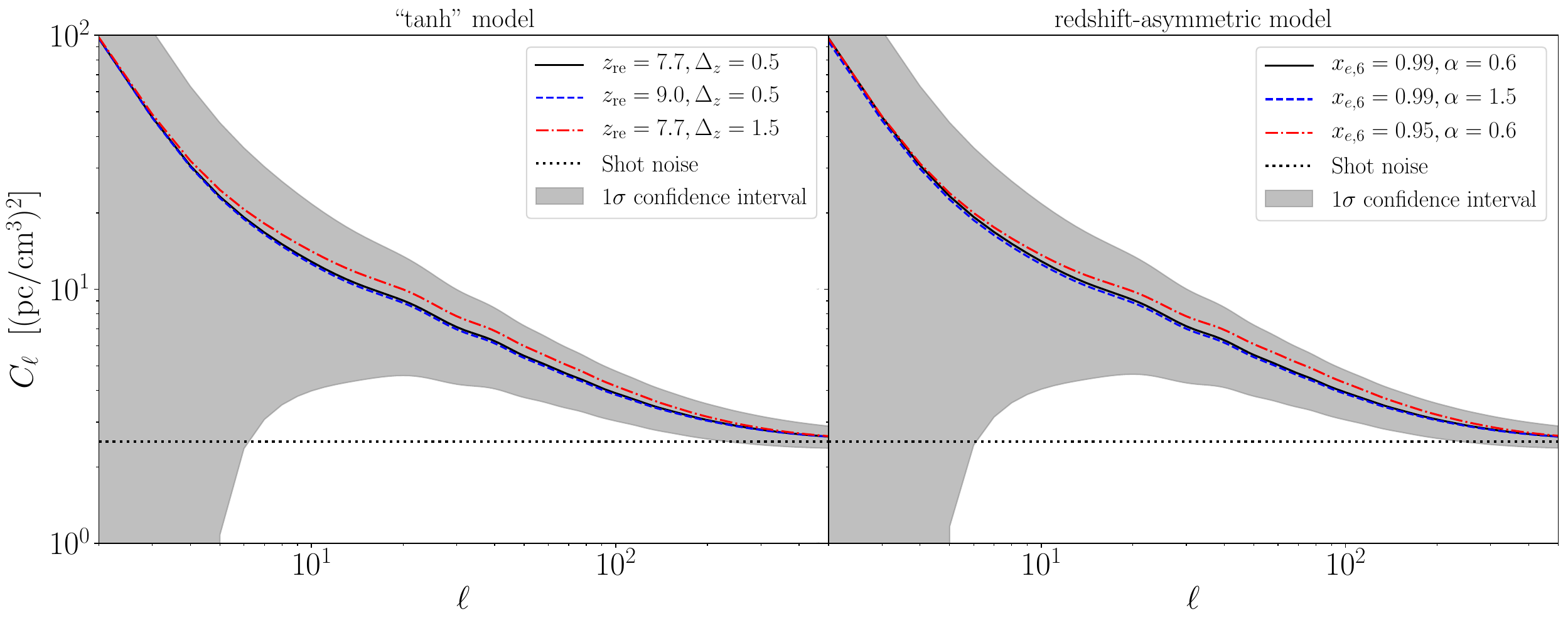}
	\caption{{The angular power spectra with different reionization parameters for the two reionization models. We also plot the $1\sigma$ confidence intervals of the fiducial models (black solid lines).}}
	\label{fig:sim}
\end{figure}

\begin{figure}[tb]
	\centering
    \includegraphics[width=1\linewidth]{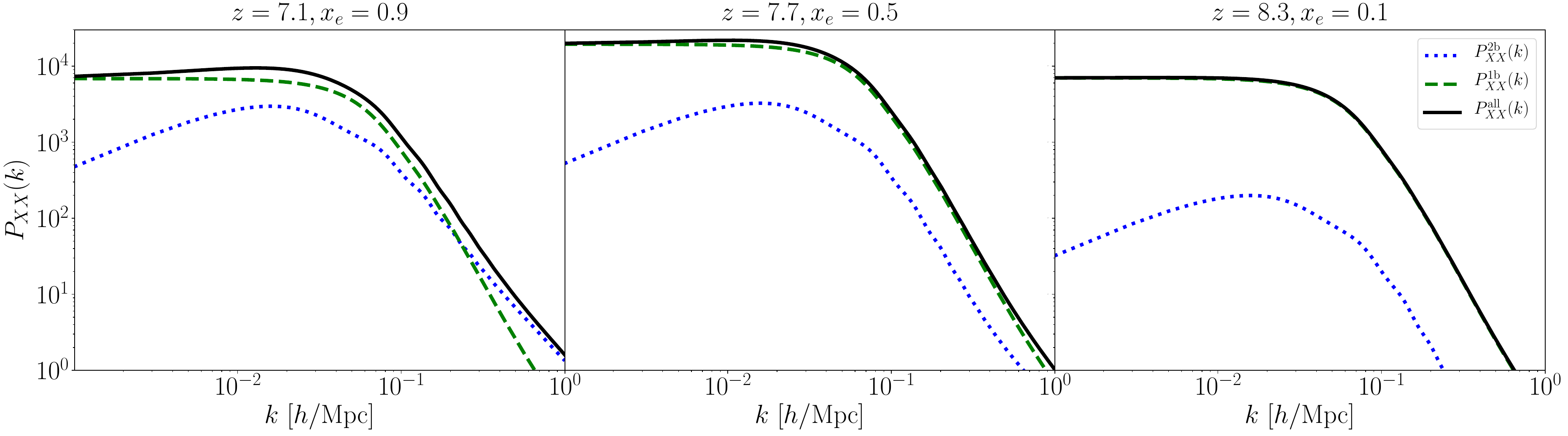}
	\caption{{$P_{XX}^{1 \mathrm{b}}(k)$ and $P_{XX}^{2 \mathrm{b}}(k)$ at different redshifts during reionization. {The green dashed lines and the blue dotted lines are for 1-bubble and 2-bubble contributions} and the black solid lines are their summations. Here we use the ``\texttt{tanh}'' model and reionization parameters are $z_{\rm re} = 7.7$ and $\Delta_z = 0.5$.}}
	\label{fig:pkxx}
\end{figure}

\begin{figure}[tb]
	\centering
    \includegraphics[width=0.6\linewidth]{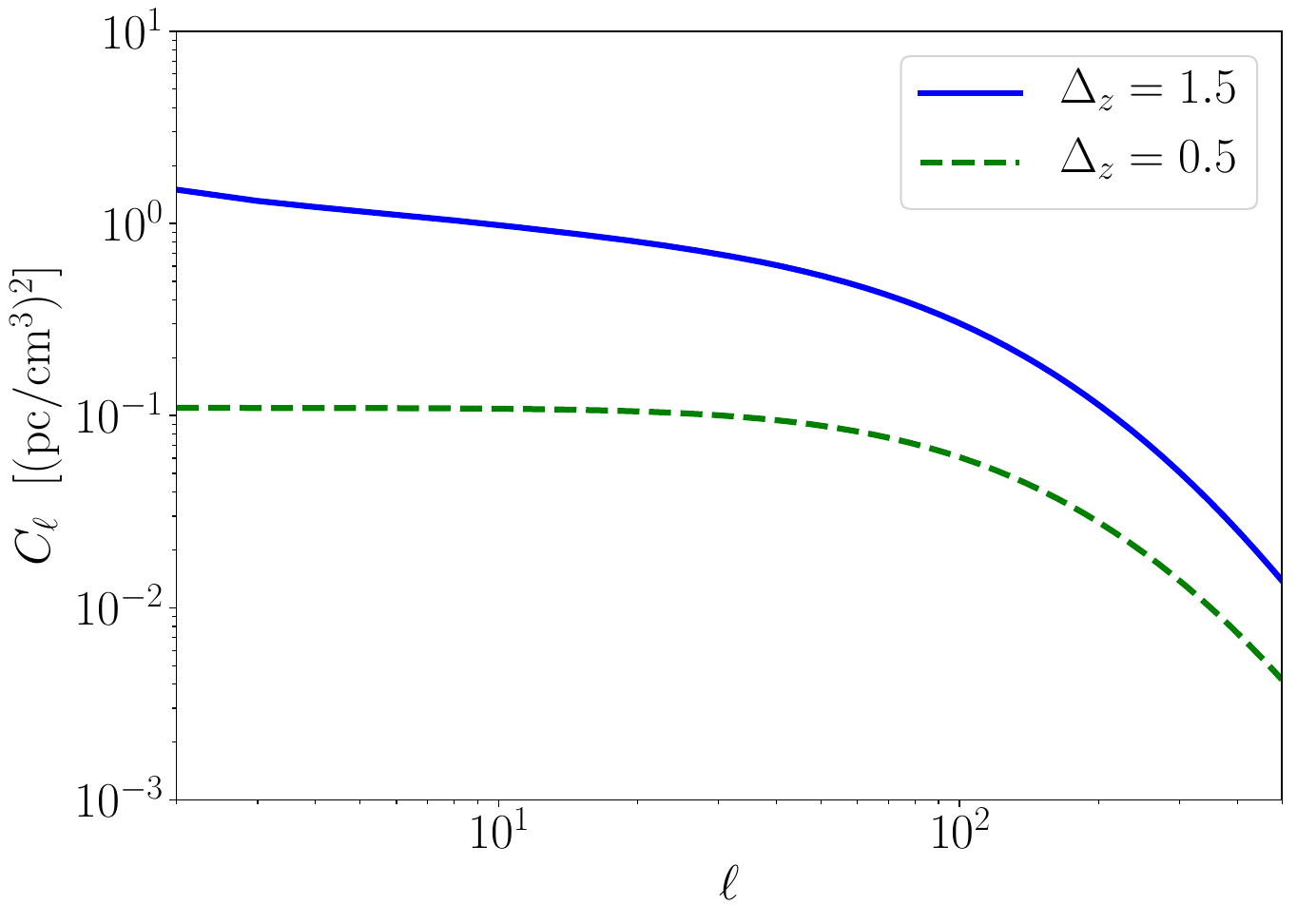}
	\caption{{The 1-bubble contributions to the angular power spectra with different $\Delta_z$. Here we set $z_{\rm re}=7.7$}.}
	\label{fig:cl1b}
\end{figure}

Then we need to consider the noise spectrum for observed DM of FRBs, which can be decomposed as \citep{Shirasaki:2017otr}
\be
\label{eq:nps}{
N^{\DM,\DM}_{\ell} = \sqrt{\frac{2}{(2\ell+1)f_{\rm sky}}} \left[C^{\DM,\DM}_{\ell}+N^{\DM,\DM}_{\ell,\rm SN}\right]~}.
\ee
{In Fig. \ref{fig:sim} we show the theoretical spectra using different reionization parameters together with the $1\sigma$ confidence intervals of the fiducial models (black solid lines).}
We can find for the ``\texttt{tanh}" model, the APS is enhanced when $\Delta_z$ gets larger, that is, with a wider reionization duration. The reason can be found in Eq.(\ref{eq:pk1b}) which shows  $P_{XX}^{1 \mathrm{b}}(k) = 0$ whether $x_e=0$ or $x_e=1$. {Besides, the  1-bubble term dominates $P_{XX}(k)$ during reionization as Fig. \ref{fig:pkxx} shows, so a wider reionization duration can induce a larger APS. Here we must emphasize that although $P_{XX}^{1 \mathrm{b}}(k)$ is dominated during reionization, its  final contribution  to the angular power spectrum is not significant since $P_{XX}^{1 \mathrm{b}}(k) = 0$ at lower redshifts where  window function $W_{{\DM},{\IGM}}(z)$ is larger. Fortunately, the 1-bubble contribution to APS is very sensitive to the reionization duration, and $C_{\ell}^{\rm 1b}$ is 10 times larger when $\Delta_z=1.5$ comparing with $\Delta_z=0.5$ as Fig. \ref{fig:cl1b} shows. So the hump of the red line in the left panel of Fig. 2 is mainly induced by $C_{\ell}^{\rm 1b}$.}
{Besides, we can find DM APS is hardly changed with a larger $z_{\rm re}$ because $W_{{\DM},{\IGM}}(z)$ is small at larger redshifts.} However, since CMB is sensitive to $z_{\rm re}$, we can use DM spectra to constrain $\Delta_z$ as a complementary probe. As for the redshift-asymmetric model, the spectrum is enhanced with a smaller $x_{e,6}$ since $P_{XX}^{1 \mathrm{b}}(k)$ is larger at $z<6$. Besides, a larger $\alpha$ means a narrower reionization duration, thus the APS is depressed. However, DM APS is not very sensitive to this parameter since the main differences of  $P_{XX}^{1 \mathrm{b}}(k)$ caused by $\alpha$ come from higher redshifts ($z>8$).

Finally we can construct the Gaussian likelihood function, where we have assumed the different scales are independent with each other, and obtain the $\chi^2$ function,
\be
\chi^2=\left(\hat{C}_{\ell}^{\DM,\DM}-C_{\ell}^{\DM,\DM}\right) \Gamma^{-1}_{\ell,\ell'}\left(\hat{C}_{\ell'}^{\DM,\DM}-C_{\ell'}^{\DM,\DM}\right)^{\rm T},
\ee
where $C_\ell^{\DM,\DM}$ refer to the theoretical model, $\hat{C}_\ell^{\DM,\DM}$ are our mock data and $\Gamma_{\ell,\ell'} = \delta_{\ell, \ell'} (N_{\ell}^{\rm DM, \DM})^2$ is the diagonal covariance matrix. In our analysis, we set $\ell_{\max}=500$ to investigate the clustering signals from the future surveys.

\section{Results on Constraints}
\label{sec:res}
Before presenting constraints on the reionization history using different models, we need to define the beginning and the end of the reionization epoch by the redshift $z_{\rm beg}\equiv z_{10\%}$ and $z_{\rm end}\equiv z_{90\%}$ at which $x_e=0.1$ and $x_e=0.9$, and the duration of the transition, defined as $z_{\rm dur}\equiv z_{\rm beg}-z_{\rm end}$. {All the constraint results we care about are listed in Tab. \ref{tab:res}.}

\subsection{Instantaneous Model}

\begin{figure}[tb]
	\centering
	\includegraphics[width=0.6\linewidth]{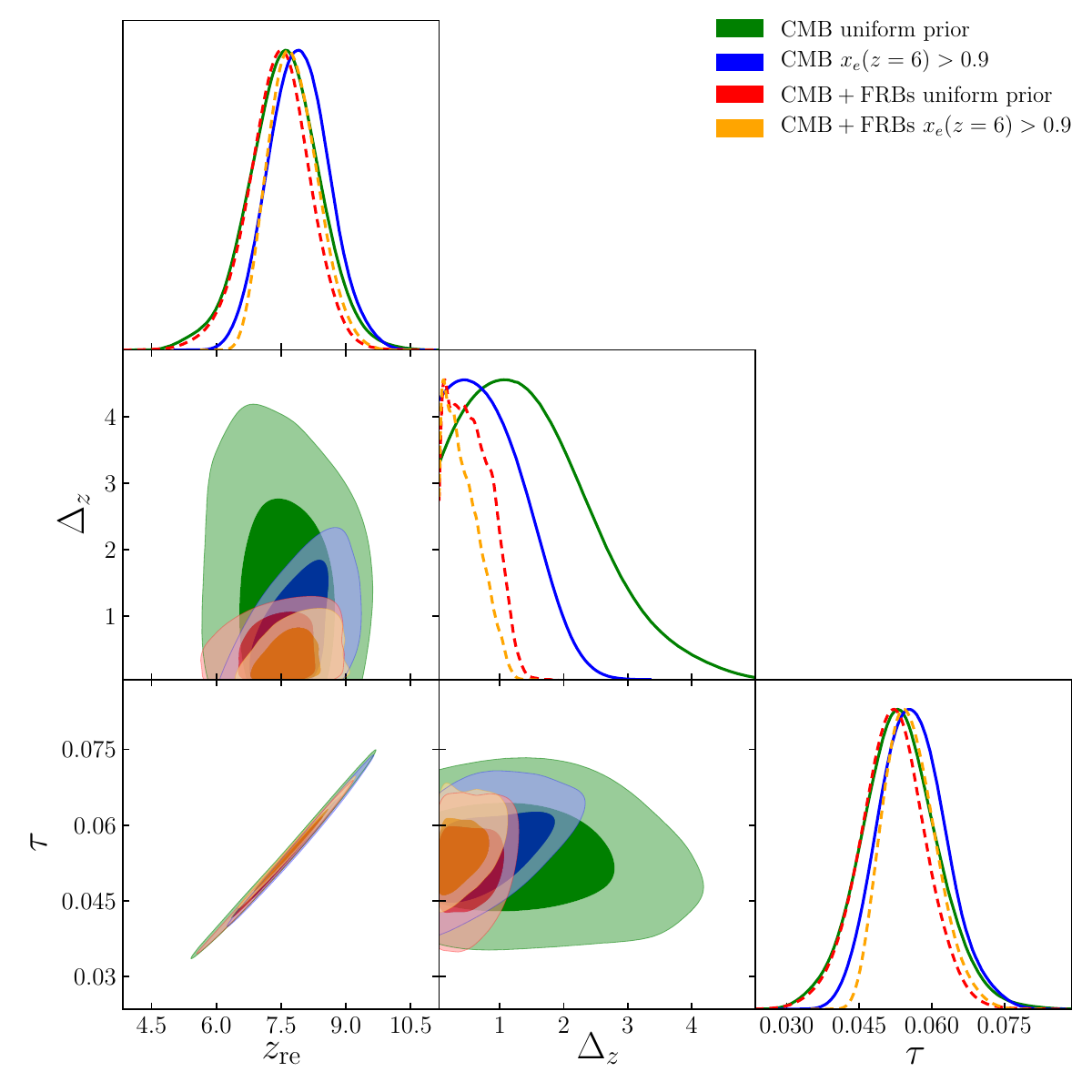}
	\caption{The marginalized one-dimensional and two-dimensional constraints on $\{z_{\rm re}, \Delta_z, \tau\}$ using mock DM angular power spectrum and Planck 2018 measurements. We use the ``\texttt{tanh}'' model and two priors here: uniform and $x_e(z=6)>0.9$.}
	\label{fig:cov2}
\end{figure}

We start with the instantaneous ``\texttt{tanh}'' model. The fiducial reionization parameters of the mock DM APS is $z_{\rm re} = 7.7$ and $\Delta_z = 0.5$, and the other parameters are fixed at the best fit values from Planck 2018 results using
TT,TE,EE+lowE. In Fig. \ref{fig:cov2} we present the $1\sigma$ and $2\sigma$ constraints of $\{z_{\rm re}, \Delta_z, \tau\}$ with two different priors. Firstly, we use the CMB measurements alone with the uniform prior, and obtain tight constraint on the redshift of reionization: {$z_{\rm re} = 7.58\pm 0.81$} at the 68\% confidence level. Consequently, the tight constraint on the optical depth is also obtained, {$\tau=0.0532\pm 0.0078$} ($1\sigma$ C.L.), since $\tau$ is mainly determined by reionization redshift $z_{\rm re}$. However, as discussed before, the CMB polarization measurements are not sensitive to the reionization width, which can only give a very weak constraint, namely the $95\%$ C.L. upper limit is {$\Delta_z<3.32$}. Furthermore, we can also compute the constraints on the beginning and the end of the reionization epoch: {$z_{\rm beg} = 9.11\pm1.19$ and $z_{\rm end}=5.96^{+1.29}_{-1.66}$} at 68\% confidence level, respectively. We can see that CMB alone can not determine the end of the reionization epoch precisely within the uniform prior. Finally, we get weak constraint on the duration of the transition: {$z_{\rm dur}<8.33$} (95\% C.L.), which means CMB alone can not verify whether the transition is instantaneous and the gentle transition is still allowed.

In order to improve the constraints, we then include the prior $x_e(z=6)<0.9$ into the calculations, which could help the CMB data to narrow the parameter space of $z_{\rm end}$. As shown in the blue contours of Fig. \ref{fig:cov2}, the constraint on $\Delta_z$ is obviously shrunk by a factor of 2, namely {$\Delta_z < 1.84$} at 95\% confidence level, while the limits of $z_{\rm re}$ and $\tau$ are only slightly tighter, {$z_{\rm re}=7.91\pm0.68$ and $\tau=0.0567\pm 0.0068$ (68\% C.L.)}. We also calculate the end of the reionization epoch and the duration of the transition, and obtain the 95\% constraints from CMB alone: {$6< z_{\rm end}<8.62$ and $z_{\rm dur}<4.25$}, respectively, which are much tighter than the uniform prior case.

\begin{table*}[tb]
	\caption{ 1$\sigma$ errors of the reionization parameters using different models and priors, while we quote 95\% upper limits for some of these parameters.}
    \label{tab:res}
    \resizebox{\textwidth}{15mm}{
    \setlength{\tabcolsep}{0.5mm}{
	\begin{tabular}{ccc ccc  ccc }
    \hline
    \hline
    \multicolumn{3}{c}{ instantaneous model, uniform prior} &\multicolumn{3}{c}{ instantaneous model ($x_e(6)>0.9$)} & \multicolumn{3}{c}{Redshift-asymmetric model ($x_e(6)>0.9$)} \\
    \hline
    Parameter & CMB  & CMB+FRBs &Parameter & CMB  & CMB+FRBs & Parameter & CMB  & CMB+FRBs  \\
    \hline
    $z_{\rm re}$ & $7.58\pm0.81$  &  $7.47\pm0.72$ & $z_{\rm re}$ & $7.91\pm0.68$  &  $7.75\pm 0.58$ & $x_{e,6}$ & $0.950\pm0.034$  &  $0.985\pm0.012$  \\
    $\Delta_z$  & $<3.32$  &  $<1.18$  &$\Delta_z$  & $<1.84$  &  $<1.02$  & $\alpha$  & $0.716^{+0.289}_{-0.212}$  &  $0.773^{+0.258}_{-0.223}$ \\
    $\tau$    &  $0.0532\pm 0.0078$ & $0.0519\pm 0.0067$ & $\tau$    &  $0.0567\pm 0.0068$  &  $0.0551\pm 0.0055$ &$\tau$    &  $0.0531^{+0.0076}_{-0.0049}$  &  $0.0525^{+0.0067}_{-0.0038}$   \\
    $z_{\rm beg}$  &  $9.11\pm1.19$  &  $8.15\pm0.95$ &$z_{\rm beg}$  &  $8.96\pm1.09$  &  $8.32\pm0.77$ &$z_{\rm beg}$ &  $<11.89$  &  $<11.32$  \\
    $z_{\rm end}$ &  $5.96^{+1.29}_{-1.66}$  &   $6.98\pm0.77$  & $z_{\rm end}$ &  $<8.62$  &   $7.37\pm0.58$   &$z_{\rm end}$  &  $<6.22$  &   $6.143^{+0.048}_{-0.041}$  \\
    $z_{\rm dur}$ &  $<8.33$   &   $<2.59$  & $z_{\rm dur}$ &  $<4.25$   &   $<2.24$  &$z_{\rm dur}$  &  $<5.64$   &   $<5.08$ \\
    \hline
	\end{tabular}}}
    \centering
\end{table*}

Next, we combine the CMB data and the mock DM power spectrum information together to reconstruct the reionization history. We still start with the uniform prior. Due to the constraining power of DM power spectrum, the 95\% upper limit of width is significantly shrunk by { a factor of 3, $\Delta_z < 1.18$}. Consequently, the constraint on the end of the reionization epoch also becomes tighter, namely {$z_{\rm end}=6.98\pm0.77$} (68\% C.L.). On the other hand, since this DM measurement is not sensitive to the $z_{\rm re}$, the constraints on $z_{\rm re}$ and $\tau$ at $1\sigma$ confidence level are only slightly improved, {$z_{\rm re}=7.47\pm0.72$ and $\tau=0.0519\pm 0.0067$}, within the uniform prior. Furthermore, since the mock FRBs sample does not have too much sources at $z>7.5$, the constraint on $z_{\rm beg}$ is only tighter about {20\%}, {$z_{\rm beg}=8.15\pm0.95$} (68\% C.L.). Finally, we obtain the constraint on the duration time, {$z_{\rm dur}<2.59$ at 95\% confidence level, which is 3.2 times tighter} than that from CMB data alone.

Finally, we include the prior $x_e(z=6)<0.9$, and find that the constraints become much tighter further. The constraining power of DM power spectrum is still useful and narrow the limit of the width to {$\Delta_z < 1.02$} at 95\% confidence level, which is still improved by a factor of 2 when comparing with the constraint from CMB data alone. Similarly, the determination of the end of the reionization epoch becomes preciser, namely {$z_{\rm end}=7.37\pm0.58$} (68\% C.L.). In the meanwhile, the $1\sigma$ constraints on $z_{\rm re}$, $\tau$ and $z_{\rm beg}$ are also slightly improved, {$z_{\rm re}=7.75\pm0.58$, $\tau=0.0551\pm 0.0055$ and $z_{\rm beg}=8.32\pm0.77$}, respectively. Again, we get the final constraint on the duration of the transition, {$z_{\rm dur}<2.24$} at 95\% confidence level. Based on these results, we can see that, different from the CMB data, the DM information is very sensitive to the transition shape of the reionization epoch. The constraining power of DM power spectrum could be a helpful complementary measurement which significantly improves our understanding on the reionization history.

\subsection{Redshift-asymmetric Model}

\begin{figure}[tb]
	\centering
	\includegraphics[width=0.6\linewidth]{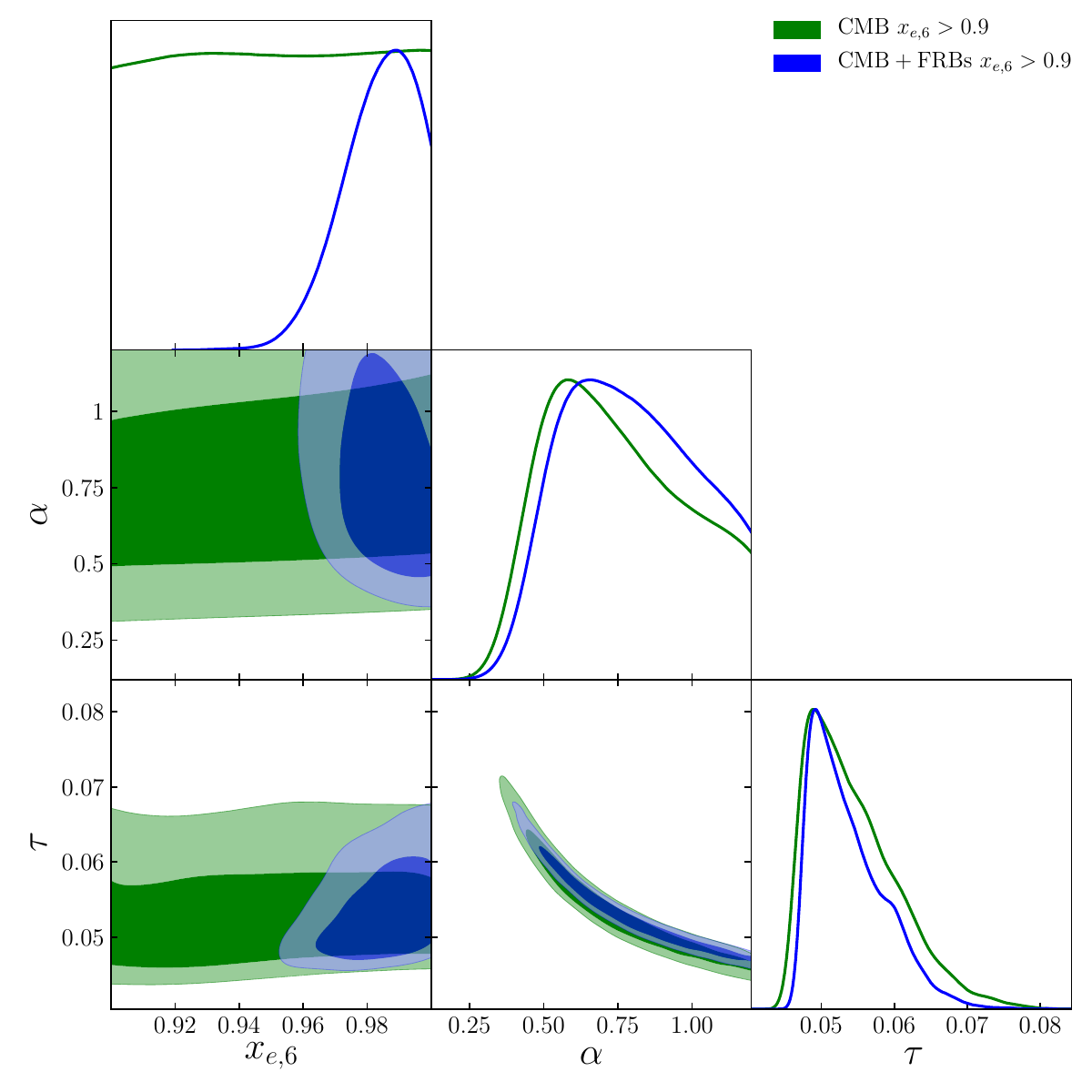}
	\caption{The marginalized one-dimensional and two-dimensional contours on $\{x_{e,6}, \alpha, \tau \}$ using CMB and FRBs measurements. Here we use the redshift-asymmetric model and the $x_{e,6}>0.9$ prior.}
	\label{fig:covas}
\end{figure}

Finally let us move to the redshift-asymmetric model, in which we only consider the prior $x_{e,6}>0.9$. We generate the mock DM angular power spectrum using $x_{e,6}=0.99$, $\alpha=0.6$, and the constraint results are shown in Fig. \ref{fig:covas}. When we use the CMB measurements alone with the $x_{e,6}>0.9$ prior, we can clearly see that the ionization fraction at $z=6$, $x_{e,6}$, cannot be constrained, since the CMB data is only sensitive to $\tau$. The marginalized $1\sigma$ constraints on the exponent $\alpha$ and the optical depth $\tau$ are: {$\alpha=0.716^{+0.289}_{-0.212}$ and $\tau=0.0531^{+0.0076}_{-0.0049}$}. We can also calculate the beginning and the end of the reionization epoch, and finally obtain the duration of the transition: {$z_{\rm dur}<5.64$} at 95\% confidence level. When we add the mock DM power spectrum into calculation, the strong constraining power will immediately shrink the constraint of the ionization fraction at $z=6$ to {$x_{e,6}=0.985\pm0.012$} at 68\% confidence level. This means the DM measurement could be very sensitive to the ionization fraction after the reionization transition time, which could be very useful to check whether our Universe is fully reionized at some redshift. { As for $\alpha$ and $\tau$, there are not significant improvements, which are {$\alpha=0.773^{+0.258}_{-0.223}$ and $\tau=0.0525^{+0.0067}_{-0.0038}$} ($1\sigma$ CL.). In this model, the end of the transition is strictly constrained, and the beginning of the transition is mainly determined by $\alpha$, so the duration of the epoch of reionization is only slightly shrunk: {$z_{\rm dur}<5.08 (95\%~\rm CL)$.}}

\section{Conclusions}
\label{sec:con}
In this paper we firstly use the auto-correlation power spectrum of DM measurement from mock FRB catalog to constrain the reionization history, which only need the rough redshift distribution of the catalog, instead of the precise redshift information of each source. In the instantaneous ``\texttt{tanh}'' model, different from the CMB data which can only constrain the reionization redshift $z_{\rm re}$, the DM measurement from about $10^4$ FRBs {spanning 20\% of all sky} can provide very useful information on the transition shape of the reionization epoch, and significantly improve the constraints of the transition width $\Delta_z$ and the duration time $z_{\rm dur}$. We also check the redshift-asymmetric model and find that the DM measurement is very sensitive to the ionization fraction after the reionization transition time, which can not be done by the CMB data alone.

{Since the shot noise is proportional to $\sigma^2_{\rm host}/N$, smaller samples will induce a larger shot noise. We checked the results for $N=5000$ with ``\texttt{tanh}" model and uniform prior. The constraint is {$\Delta_z < 1.21$ (95\% C.L.) and the marginalized error is increased by 18.6\%}. Actually, Recent works (e.g. \citep{Hashimoto:2020dud}) show FRBs will be detected with SKA at a rate of $\sim 10^3-10^4~\rm (sky^{-1} day^{-1})$ and $\sigma_{\rm host} = 100 \rm pc/cm^3$ used in our work is very conservative. The signal to noise ratio will be much improved with the future surveys. Thus, this method could be a helpful complementary measurement which can significantly improve our understanding on the reionization history.
}

\section*{Acknowledgements}

We thank Z.-X. Li and H. Gao for useful discussions. This work is supported by the National Science Foundation
of China under grants No. U1931202 and 12021003, and the National Key R\&D Program of China under
grant No. 2017YFA0402600.


\providecommand{\href}[2]{#2}\begingroup\raggedright\endgroup

\end{document}